\documentstyle[preprint,prl,aps]{revtex}

\begin{document}
\preprint{\vbox{\hbox{October 1995}\hbox{IFP-715-UNC}}}
\draft
\title{Dicyclic Horizontal Symmetry and Supersymmetric Grand Unification.}
\author{\bf Paul H. Frampton and Otto C. W. Kong}
\address{Institute of Field Physics, Department of Physics and Astronomy,\\
University of North Carolina, Chapel Hill, NC  27599-3255}
\maketitle

\begin{abstract}
It is shown how to use as horizontal symmetry the dicyclic group $Q_6
\subset SU(2)$ in a supersymmetric unification $SU(5)\otimes SU(5)\otimes
SU(2)$ where
one $SU(5)$ acts on the first and second families, in a
horizontal doublet, and the other acts on the third. This can lead to
acceptable quark masses and mixings, with an economic choice of matter
supermultiplets, and charged lepton masses can be accommodated.

\end{abstract}

\newpage

The smallness of most of the quark masses and mixing parameters and the
strong hierarchy among them is one of the most interesting puzzles in particle
physics. Spontaneously broken horizontal symmetry is the most popular
candidate theory for  understanding the flavor structure,
including in supersymmetric models. In the context of
the MSSM,  a horizontal symmetry may also give a viable
alternative to build in
a super-GIM mechanism  to suppress FCNC induced by
supersymmetric particles\cite{SD,QSA,PT}. Attempts has also  been made
to use  horizontal symmetry to address the $\mu$-problem\cite{QSA,mu}, the
strong CP problem\cite{cp}, FCNC due to light leptoquarks\cite{lq}, and
baryon number violation in supersymmetry\cite{bnv}. There is hence a growing
interest in the topic.

However, as global  symmetries
 are in general not respected by gravitational effects\cite{gg},
the horizontal symmetry should be gauged. Canceling the gauge
anomalies then imposes a strong constraint on model
building\cite{U1g,DPS,su2,q2n2}.
For a simple nonabelian symmetry, we are left with essentially only $SU(2)$
and its discrete dicyclic subgroups $Q_{2N}$\cite{q2n2,q2n,gna,HM}.

	Now we consider an extra desirable ingredient, compatibility with
supersymmetric vertical (grand) unification, like $SU(5)$. The only
 GUT-compatible gauged horizontal symmetry model
proposed so far is incompatible with SUSY\cite{q2n2}.
Here  we provide the first SUSY-GUT compatible such model.

	Inspired by the anti-unification approach to quark masses\cite{ATG},
models with separate GUT groups for each of the three families
 has been introduced\cite{555}.
Here we consider instead only two $SU(5)$'s for  horizontal singlet and
doublet families. The structure then gives, to the first approximation,
 rank one  quark mass matrices. We show that, with  judiciously chosen
 heavy scalar VEVs, the full hierarchical and phenomenologically-viable
quark mass matrix textures can be generated, using
nonrenormalizable gravitational interactions\cite{EG}.

Our model has gauged $SU(5)\otimes SU(5)\otimes SU(2)$, with this symmetry
broken to a diagonal $SU(5)$ (SUSY-)GUT group around and above the GUT
scale. The full pattern of symmetry breaking is illustrated in Figure 1.

The assignment of the three families of quarks and leptons to
$(SU(5)\otimes SU(5)\otimes Q_6)$ is thus
\begin{tabbing}
aaaaaaaaaaaaaaaaa\= for 1st and 2nd families\= $(\bar{5}+10, 1, 1)$ \kill
 	\> 3rd family  \> $(\bar{5}+10, 1, 1)$ 		\\
	\> 1st and 2nd families \> $(1, \bar{5}+10, 2_1)$.
\end{tabbing}
Upon breaking to diagonal $SU(5)$ this becomes a normal 3-family SUSY-GUT.

The Higgses which will break electroweak symmetry are in
$(5+\bar{5},1,1)$ and so couple only to the third family in a renormalizable
fashion. Scalar VEVs in $(\bar{5},5)$ or $(5,\bar{5})$ will break
 to the diagonal subgroup. There will also be
$SU(5)\otimes SU(5)$ singlets, non-trivial under $Q_6$. Beyond these
scalars, it will be necessary only to introduce an extra $(15,\bar{10},2_1)$
multiplet to complete the model.

Taking as an expansion parameter  $\lambda \sim sin\theta _c \sim .22$
we will use two scale below $M_{Planck}$ which are taken as
$M_{1} \sim \lambda \tilde{M}_{Planck}$ which characterizes the VEV of a
$(1,1,2_1)$ and $M_{2} \sim \lambda ^3 \tilde{M}_{Planck}$ which sets the
$SU(5)\otimes SU(5)$ breaking VEVs. In fact, $M_{2}$ lies just above the
usual $M_{GUT} \approx 2\times 10^{16} GeV$, as the effective Planck mass
$\tilde{M}_{Planck}$ is given by $M_{Planck} /
 \sqrt{8 \pi } \approx 2.4\times 10^{18} GeV$.
Thus, the hierarchy of the observed quark masses at accessible energy
merely reflect the existence of the superheavy scalars lying at and above
$M_{GUT}$.

To keep track of the book-keeping for the components of the $Q_6$ couplings,
we find it most convenient to assign to the two components of a $2_n$
doublet the values $\pm n$, reflecting the eigenvalues of $2T_3 = \pm n$
in the natural embedding $SU(2)\supset Q_{2N}$. Recall that for
even-dimensional
$SU(2)$ irreducible representations
\[ d \longrightarrow 2_1 + 2_3 + .... \ \ \ 2_{d-1} \]
while for odd $d$,
 \[ d \longrightarrow singlet + 2_2 + 2_4 + .... \ \ \ 2_{d-1}; \]
the singlet is $1$ for $d = 1, 5, 9, .... \ \ $ and $1^{'}$
for $d = 3, 7, 11, .... \ \ $. Of course we will need only $d = 1, 2, 3$ and
$4$ for $Q_6$.

With this book-keeping, we find that the mass matrix textures
emerge from VEVs as follows:
\begin{itemize}
\item the only scalar with VEV at scale $M_1$ is $(1, 1, [+1])$
\item the VEVs at scale $M_2$ are
\begin{itemize}
\item $(\bar{5}, 5, [-3])$
\item $(5, \bar{5}, [-2])$ and $(5, \bar{5}, [-1])$
\item $(15, \bar{10}, [-1])$
\end{itemize}
\end{itemize}
where the $Q_6$ entry implies the $2T_3$ eigenvalue.
Tracking down all the entries of the mass matrices to the lowest order in
$\lambda$, we have the follwing result:
\begin{equation}
$$M_{u} \sim \left(\begin{array}{ccc}
\lambda^{8} & \lambda^{6} & \lambda^{9} \\
\lambda^{6} & \lambda^{4} & \lambda^{7} \\
\lambda^{9} & \lambda^{7} & 1
\end{array}\right)$$
\end{equation}
\begin{equation}
$$M_{d} \sim \left(\begin{array}{ccc}
\lambda^{6} & \lambda^{4} & \lambda^{5}  \\
\lambda^{4} & \lambda^{3} & \lambda^{3} \\
\lambda^{7} & \lambda^{5} & 1
\end{array}\right)$$
\end{equation}

The authors of \cite{RRR} have analyzed all possible symmetric quark
 mass matrices with the maximal (six) and next-to-maximal (five)
 number of texture zeros, and concluded that only five models,
denoted by the roman numerals I to V in their work,
are phenomenologically viable. Note that the symmetric structure is just an
input assumption. In our case, the GUT structure enforced a symmetric mass
matrix for the up-sector, but leaves that for the down-sector arbitrary.
While $U(1)$ flavor symmetry constructions for quark mass matrices
with  nonsymmetric hierarchical textures have  been
attempted\cite{QSA,DPS,LNS},
the full list of such phenomenologically viable quark mass matrices
is not yet available. However, we can simply exploit the fact that
low-energy physics is unaffected by an arbitrary rotation of the right-handed
quark fields. Discarding the large order entries and imposing a rotation on the
right-handed second and third down-quark fields with an angle $\sim
\lambda^{3}$, we
obtain, in the symmetric basis,
\begin{equation}
$$M_{u} \sim \left(\begin{array}{ccc}
0 & \lambda^{6} & 0 \\
\lambda^{6} & \lambda^{4} & 0 \\
0 & 0 & 1
\end{array}\right)$$
\end{equation}
\begin{equation}
$$M_{d} \sim \left(\begin{array}{ccc}
0 & \lambda^{4} & 0  \\
\lambda^{4} & \lambda^{3} & \lambda^{3} \\
0 & \lambda^{3} & 1
\end{array}\right)$$
\end{equation}
corresponding to case I of ref.~\cite{RRR}, hence showing that
 the (asymmetric) quark mass texture is phenomenologically
viable.

Now we turn to the charged lepton mass matrix. The simplest way to accommodate
it  to obtain the Georgi-Jarlskog pattern\cite{GJ} by replacing the scalar
VEV  $(5,\bar{5},[-2])$, which is responsible for the $(M_d)_{22}$ entry,
 with a  $(5,\bar{45},[-2])$. If one wants to avoid having a $\bar{45}$,
there is the alternative suggested by Ellis and Gaillard\cite{EG}. While
the diagonal $SU(5)$ singlet from the $(5,\bar{5})$ contributions to
the quark and lepton masses are the same, the other $SU(3) \otimes SU(2)
\otimes U(1)$
singlet in the adjoint $24$ of the diagonal $SU(5)$ gives quark and lepton
masses in
the ratio $-3/2$. Ellis and Gaillard showed that if both the singlet and
the $24$ contribute, with partial cancellation in the lepton-sector, the
$\bar{5}$ VEV could fit both the quark- and lepton-sectors. The $24$ VEV is
of course GUT-breaking, which is needed anyway. In our case, its
 contribution has to be smaller by
about a factor of $M_{GUT}/M_{2}$. Without going into detail, a simple
comparison with the Ellis-Gaillard analysis shows that
this is can be successful.

Having discussed both the quark and charged-lepton mass matrix texture
construction, a few comments are in order:
\begin{itemize}
\item The breaking of the horizontal $SU(2)$ through the discrete dicyclic
subgroup $Q_6$ is needed to avoid the otherwise large D-term contributions
to the scalar quark masses in the $SU(2)$ breaking\cite{PT}. Our model may
otherwise be considered only in the $SU(2)$ framework. However, the D-term
contributions  would lift any
assumed degeneracy among the squarks and cause unacceptable FCNC in
for example $K-\bar{K}$ mixing (see \cite{HKT} and references therein).
 The  strongest FCNC constraint can
be expressed as an upper limit on the (12) entry of the left-handed
down-squark $\tilde{m}^2 _{LL}$ matrix, in the quark-mass eigenbasis
\begin{equation}
\delta \tilde{m}^2 _{ds} = \tilde{m}^2 _1 K_{11} K^{\dag}_{12} +  \tilde{m}^2
_2
 K_{12}  K^{\dag}_{22} + \tilde{m}^2 _3  K_{13}  K^{\dag}_{32}
\end{equation}
where $\tilde{m}^2 _i$ are the three eigenvalues
and $K$ the unitary transformation matrix that diagonalizes $\tilde{m}^2
_{LL}$. In the limit that $K_{13}  K^{\dag}_{32}$ is negligible, this reduces
to
\begin{equation}
\delta \tilde{m}^2 _{ds} \approx (\tilde{m}^2 _2 -  \tilde{m}^2 _1) K_{12}
\end{equation}
hence a degeneracy condition  between $\tilde{m}^2 _1$ and $ \tilde{m}^2 _2$,
unless the mixing $K_{12}$ is itself exceedingly small\cite{qsa}. As noted in
ref.\cite{PT},
the $2+1$ family structure, gives a natural first order
degeneracy between $\tilde{m}^2 _1$ and $ \tilde{m}^2 _2$, and is therefore
flavorable from the perspective. The degeneracy is however lifted as the
horizontal symmetry is broken. In our model, the lifting is of order $\lambda
^{2}$ which is too large. Ex
tra mechanisms, as proposed in ref.~\cite{PT}, is needed to help suppress the
FCNC.
\item In principle, the non-renormalizable mass terms may be obtained,
alternatively, from the Froggatt-Nielsen mechanism\cite{FN}. In that case,
one needs $M_1/M_0 \sim \lambda$ and $M_2/M_0 \sim \lambda ^3$ where $M_0$
is the mass scale of the vector-like fermions mediating the Yukawa vertices
involving the chiral fermions. However, $M_0$ cannot really be brought
down much below $~{M}_{Planck}$ because the proliferation of heavy
supermultiplets  may lead to a non-perturbative gauge coupling\cite{QSA}.
\item The supermultiplets that contain the $SU(5) \otimes SU(5)$ breaking
VEVs in the model are assumed to be vector-like. Hence they are heavy and
have no contribution to gauge anomalies. The supermultiplet $(1,1,2_1)$ can
have heavy Majorana mass. It has a contribution which helps to cancel the
otherwise non-trivial global-$SU(2)$ anomaly. Local gauge anomaly cancellation
in our model is completely straightforward with no additional states.
\item The $(1,1,2_1)$ can be identified naturally as a right-handed
neutrino supermultiplet. If an extra $(1,1,1)$ is added, the family structure
of the right-handed neutrinos is then the same as the quarks and leptons.
While this appears natural, the neutrino masses and mixings hence derived,
assuming no extra VEVs, do not look very good. However, the
right-handed-neutrino-sector need not have the same family structure  as
the quarks and leptons, so long as the global-$SU(2)$ anomaly condition is
satisfied; and there could be some extra multiplets with or without VEVs among
them that modify  the neutrino masses and mixings without upsetting
the quark and chagred-lepton mass textures.
\item Like the minimal SUSY-$SU(5)$, the model has dimension-4 baryon number
violating operators that have to be removed by imposing R-parity or otherwise.
In particular the $(\bar{5},1,1)$ Higgs supermultiplet definitely has to be
distinguished from the quark-lepton one in the same representations to avoid
fast proton decay. Finally, the infamous doublet-triplet splitting problem has
not been addressed.

\end{itemize}

This work was supported in part by the U.S. Department of
Energy under Grant DE-FG05-85ER-40219, Task B.\\

\bigskip
\bigskip

{\bf Figure Caption.}\\

Fig.1 Illustration of the symmetry breaking pattern of the model.

\newpage

\begin{figure}

\setlength{\unitlength}{1.0cm}
\begin{center}

\begin{picture}(11,11)
\put(1.1,10){\framebox(4.8,0.7){$SU(5)\otimes SU(5)\otimes SU(2)$}}
\put(1.5,8){\framebox(4.0,0.7){$SU(5)\otimes SU(5)\otimes Q_{6}$}}
\put(2,6){\framebox(3,0.7){$SU(5)\otimes SU(5)$}}
\put(2.5,4){\framebox(1.6,0.7){$SU(5)_D$}}
\put(1,1.5){\framebox(5,0.7){$SU(3)_c\otimes SU(2)_L\otimes U(1)_Y$}}
\put(7,8){$M_1 \sim \lambda \tilde{M}_{Planck}$}
\put(7,6){$M_2 \sim \lambda ^3 \tilde{M}_{Planck}$}
\put(7,4){$M_{GUT}$}
\put(7,2.5){$M_{SUSY}$}
\put(7,1.5){$M_W$}
\put(3.3,9.8){\vector(0,-1){0.9}}
\put(3.3,7.8){\vector(0,-1){0.9}}
\put(3.3,5.8){\vector(0,-1){0.9}}
\put(3.3,3.8){\vector(0,-1){1.4}}
\put(3.3,1.3){\vector(0,-1){0.9}}

\end{picture}
\end{center}

\caption{Illustration of the symmetry breaking pattern of the model}
\label{Fig. 1}

\end{figure}
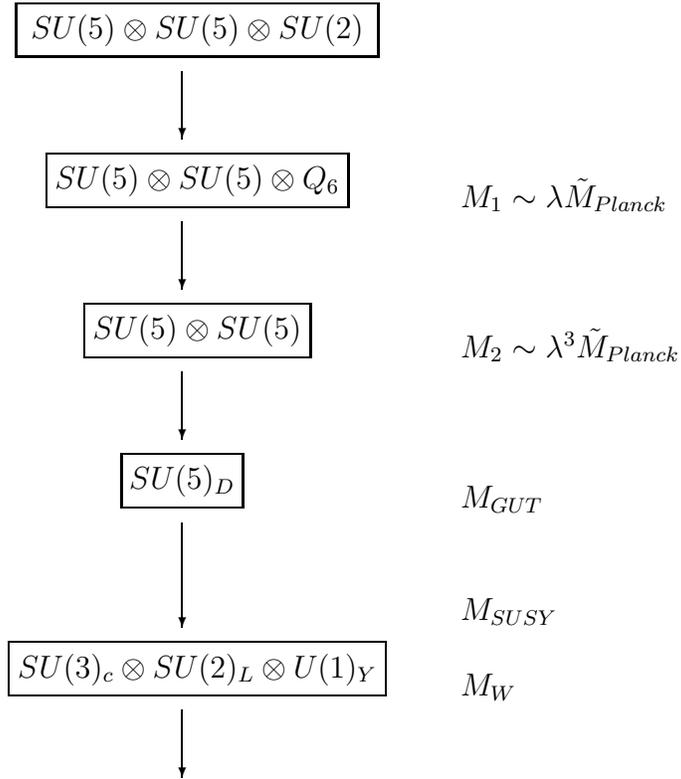


\begin{references}
\bibitem{qsa}
This is the quark-squark alignment scenario suggested in ref.~\cite{QSA}.
\end{references}
\end{document}